# Semi-flexible bimetal-based thermal energy harvesters

**S Boisseau[1], G. Despesse[1], S. Monfray[2], O. Puscasu[2], T. Skotnicki[2]**

[1] CEA, Leti, Minatec Campus, 17 rue des Martyrs - 38054 Grenoble Cedex 9, France
[2] STMicroelectronics, 850 rue J. Monnet - 38926 Crolles, France

E-mail: sebastien.boisseau@cea.fr

**Abstract.** This paper introduces a new semi-flexible device able to turn thermal gradients into electricity by using a curved bimetal coupled to an electret-based converter. In fact, a two-steps conversion is carried out: (i) a curved bimetal turns the thermal gradient into a mechanical oscillation that is then (ii) converted into electricity thanks to an electrostatic converter using electrets in Teflon®. The semi-flexible and low cost design of these new energy converters pave the way to mass production over large areas of thermal energy harvesters. Raw output powers up to 13.46µW per device were reached on a hot source at 60°C and forced convection. Then, a DC-to-DC flyback converter has been sized to turn the energy harvesters' raw output powers into a viable supply source for an electronic circuit (DC@3V). At the end, 10µW of directly usable output power were reached with 3 devices, which is compatible with Wireless Sensor Networks powering applications.

## 1. Introduction

Energy harvesting (EH) is a field of growing interest with a market expected to reach some billions of dollars within few years [1]. Many principles of ambient energy harvesting have already been investigated [2], and among them, thermal energy harvesting from thermal gradients has proven to be particularly suitable when energy harvesting from light is not possible (inside machines, under the hood of the car, etc.), as well as vibration energy harvesting [2]. Like most of current small-scale energy harvesting concepts, thermal energy harvesting is aimed at supplying Wireless Sensor Networks to remove batteries (or at least to recharge them), giving Wireless Sensor Nodes a theoretical unlimited lifetime, and removing any maintenance issues such as battery replacement or recharging.

Generally, thermal energy harvesters are based on bimetallic junctions that generate a thermoelectric voltage (Seebeck effect) when submitted to a temperature gradient [3]. This concept is known for long and besides, some thermoelectric energy harvesters are already commercialized (MicroPelt, ThermoLife). Moreover, thermoelectric energy harvesters have quite good conversion efficiencies that may reach up to 10-15% of the Carnot cycle efficiency. Yet, they require quite expensive materials such as bismuth telluride, and developing flexible devices is rather complicated.

Here, an alternative to these standard thermoelectric devices is proposed. Thermal gradients are turned into electricity by a two-steps conversion: (i) the thermal gradient is turned into a mechanical movement thanks to a curved bimetal, which is able to snap between two positions according to the temperature with a hysteresis cycle; (ii) the mechanical oscillation is then converted into electricity by using a Teflon-electret-based electrostatic converter. These new devices open the way to semi-flexible thermal energy harvesters that are compatible with mass production over large areas. After presenting the concept of bimetal-and-electret-based transducers in section 2, semi-flexible bimetal-based thermal energy harvesters are developed in section 3. Then, section 4 is focused on the Power Management Circuit, which is required to power an electronic device while optimizing power extraction from energy harvesters, validating the suitability of these thermal energy harvesters for Wireless Sensor Networks (WSN) powering applications.

## 2. Bimetal-and-electret-based converters to turn thermal gradients into electricity

Bimetals and electret-based converters are known for long, but the idea of coupling them to harvest energy from thermal gradients is quite recent [4]; this concept is presented and equated hereafter.

### 2.1 A bimetal-based heat engine

Bimetals are made of two strips of different metals with different coefficients of thermal expansion (CTE) that are joined together (e.g. iron and copper) (Figure 1(a)). CTE difference enables flat bimetallic strips to bend when heated up or cooled down, making bimetals transducers that convert temperature changes into mechanical movements. Curved or stamped bimetallic strips (Figure 1(b)) are even smarter devices, presenting strong nonlinear behaviors, and able to snap and snap-back between two positions (sudden buckling) according to the temperature with a hysteretic behavior as presented in Figure 1(c).

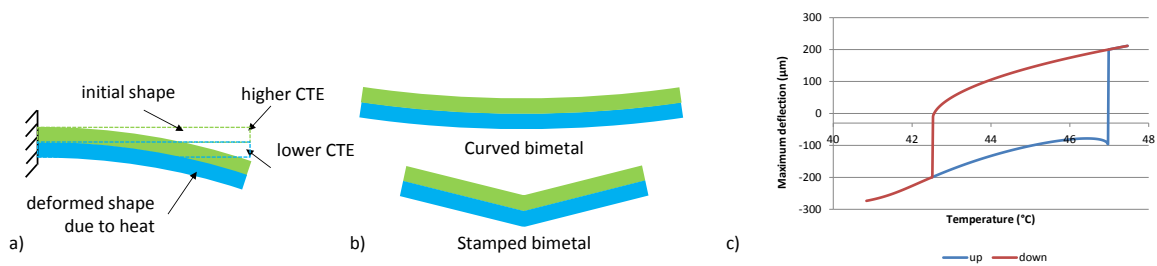

Figure 1. (a) Flat bimetals, (b) curved and stamped bimetals and (c) hysteresis cycle

Besides, this phenomenon has been thoroughly studied by Timoshenko [5] and others [6-9], and nowadays, curved and stamped bimetals are used in many electrical and mechanical devices [5, 10-11] such as actuators, clocks, thermometers, thermostats, circuit breakers, time-delay relays, etc.

In this paper, a curved bimetal is used as a heat engine capable of converting a thermal gradient into mechanical oscillations. The curved bimetal is clamped in a cavity with a hot source on the bottom and a cold source on the top (Figure 2). At the equilibrium temperature (e.g. T=25°C), the bimetal is in a convex configuration, and the metal with the higher CTE is above the metal with the lower CTE (figure 2(a)). The device is in its "lower state" and the bimetal is in contact with the lower plate.

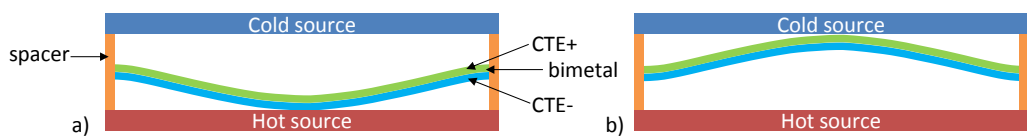

Figure 2. Bimetal-based heat engine (a) lower state and (b) upper state

Then, the device is placed on a hot source. The bimetal is heated up, accumulates mechanical elastic energy, suddenly snaps to its "upper state" (Figure 2(b)) and enters in contact with the upper plate (cold source). There, it is cooled down, snaps back to its "lower state", and a new cycle restarts.

This results in a mechanical oscillation that can be harvested thanks to a standard mechanical-to-electrical converter such as a piezoelectric, an electromagnetic or an electrostatic transducer. Actually, piezoelectric-based devices using this concept have already been proposed by [12-14].

In this paper, we focus on an electret-based electrostatic conversion to turn the bimetal's oscillations into electricity.

### 2.2 Electrets and electret-based converters

Electrets are electrically charged dielectrics that are able to keep their charges for years. They have been used for long in microphones, sensors [15], etc. and have proven their suitability in electrostatic converters as a permanent polarization source that enables a direct mechanical-to-electrical conversion without charging and discharging cycles [16-17], greatly simplifying the power management circuit.

Electret-based converters are primarily electrostatic converters and are therefore based on a capacitive architecture made of two plates (electrode and counter-electrode) as presented in Figure 3(a). The electret induces charges on electrodes and counter-electrodes to respect Gauss's law, and $Q_i$, the charge on the electret is equal to the sum of $Q_1$ and $Q_2$, where $Q_1$ is the total amount of charges on the electrode and $Q_2$ the total amount of charges on the counter-electrode ($Q_i=Q_1+Q_2$).

Then, a relative movement of the counter-electrode compared to the electret and the electrode induces a change in the capacitor geometry (e. g. the counter-electrode moves away from the electret) and leads to a reorganization of charges between the electrode and the counter-electrode through load $R$ due to charge influence variation phenomena. This results in a current circulation through $R$ and therefore, the relative movement is turned into electricity.

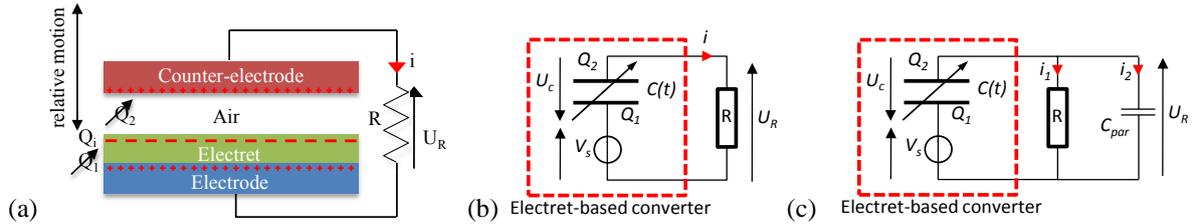

Figure 3. (a) Electret-based converter, (b) equivalent electric model with a simple load $R$ and (c) equivalent electric model taking parasitic capacitances $C_{par}$ into account [19]

The equivalent model of this converter is presented in Figure 3(b) and has already been thoroughly discussed and validated by experimental data in [19]. As a consequence, the electret-based converter is ruled by equation (1), where $V_s$ is the electret's surface voltage and $C(t)$ the capacitance of the energy harvester.

$$\frac{dQ_2}{dt} = \frac{V_s}{R} - \frac{Q_2}{C(t)R} \qquad (1)$$

And, the instantaneous output power is expressed by (2) on a simple resistive load. It is also noteworthy that this model has been validated on more complicated loads such as diode bridges, capacitors, DC-to-DC converters.

$$p(t) = R\left(\frac{dQ_2}{dt}\right)^2 \qquad (2)$$

The effect of parasitic capacitances $C_{par}$ (figure 3(c)) on the electret-based energy harvesters has also been discussed in [19]. Their impact on the energy harvester's output voltages and output powers has been in particular shown when working with high impedance loads: decrease of output voltages, strong decrease of maximum output voltages and, as a consequence, decrease of output powers.

The electret-based converter is then added to the bimetal-based heat engine to turn the mechanical oscillations into electricity.

### 2.3 Bimetal-based Thermal Energy Harvesters

Actually, the bimetal-based transducer can be easily coupled to the electret-based converter presented in Figure 3 by adding two electret layers and two electrodes between the hot and the cold plates as presented in Figure 4. In this configuration, power is harvested both on $R_1$ and $R_2$, as two electret-based converters are formed: the upper electrode, the upper electret and the bimetal for the first one and the bimetal, the lower electret and the lower electrode for the second one.

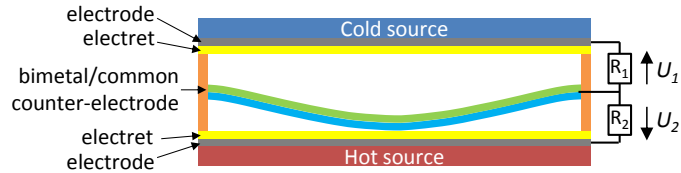

Figure 4. Bimetal and electret-based converter for thermal energy harvesting

In fact, this concept has already been validated and presented in [4] on rigid devices made of silicon plates. Here is introduced another opportunity offered by bimetal-and-electret-based thermal energy harvesters: *low-cost semi-flexible devices*.

## 3. Semi-flexible bimetal-based thermal energy harvesters

Semi-flexible (and more generally flexible) devices are a great opportunity for thermal energy harvesting, opening the door to conformable devices, adaptable to many environments such as pipes, pumps, motors, human bodies, etc.

### 3.1 Designs

The semi-flexible bimetal-based devices we propose are made from two sheets of steel (75µm) covered with a 25µm-thick Teflon® layer. A bimetal is inserted into the cavity formed by the two sheets of steel as presented in Figure 5, and oscillates when placed on a hot source between its lower and its upper state (respectively, figure 5(a) and figure 5(b)). This movement is then harvested by the two electret-based converters.

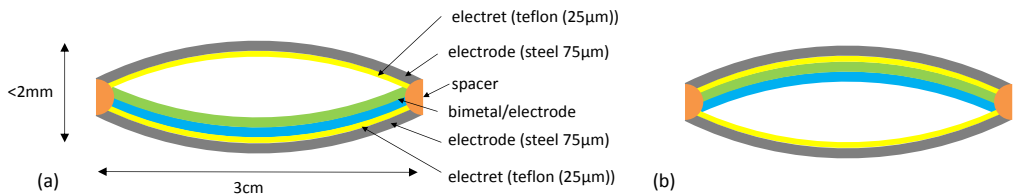

Figure 5. Semi-flexible thermal energy harvester (Teflon-based device) (a) "lower state" and (b) "upper state"

Besides being semi-flexible, such a design enables to develop low-cost systems made of simple and fully available materials such as Teflon or steel.

### 3.2 Prototypes, output voltages and output powers

The prototypes presented in this paper are made from a 115µm-thick curved bimetal designed to snap at 47°C and to snap-back at 42.5°C; but obviously, another type of bimetal could be chosen, for example to work at higher temperatures. The bimetal is in INVAR (Fe-Ni36%), which has a very low CTE ($\alpha=2\times10^{-6}$), and in B72M (Mn-Cu18%-Ni10%), which is the high CTE material ($\alpha=26.4\times10^{-6}$). The bimetal's surface is 1cm×3cm and its weight is equal to m=0.26g; it is covered by a 1µm-thick parylene-C layer to protect electrets' charges during contacts. The complete device sizes 34×12×1.5mm³ (0.6cm³).

The Teflon layers are metallized on the rear face, glued on the steel electrodes, and finally charged by a standard negative corona discharge (point-grid-plane architecture) with a point voltage of 10kV [18] during 1 hour. During the first 30 minutes, the electret is heated at 200°C on a hotplate. Then, the hotplate is turned off and the corona discharge is maintained while the electret and the hotplate cool down. This process enables to improve charge stability of Teflon electrets [19-20]. Electrets are then added around the bimetal with spacers (200µm on each side). The prototype (figure 6) is placed on a hot source at 60°C and cold by forced convection (fan); the upper plate's temperature is then $T_{cold}$=36°C. As expected, the bimetal oscillates between the two plates and the electret-based converters turn this mechanical movement into electricity. Figures 6(a, b) present a side view and a top view of the prototype and figures 6(c, d) show the two bimetal's shapes ("lower state" before the snap-through and "upper state" after the snap-through).

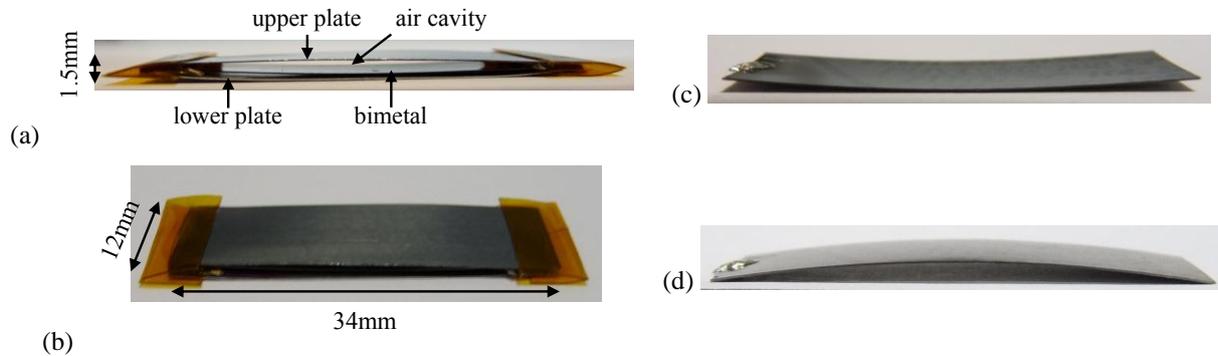

Figure 6. Prototype (a) side view (lower state) and (b) top view and bimetal's shapes (c) "lower state" and (d) "upper state"

The output voltages on a 1 GΩ load for various electret's surface voltages ($V_s$=400V, $V_s$=450V, $V_s$=500V) are presented in figure 7; the mean output powers on each channel ($R_1$ and $R_2$) are specified in the respective charts ($P_1$ for channel 1 ($R_1$), $P_2$ for channel 2 ($R_2$) and $P= P_1+P_2$, the total mean output power of the device).

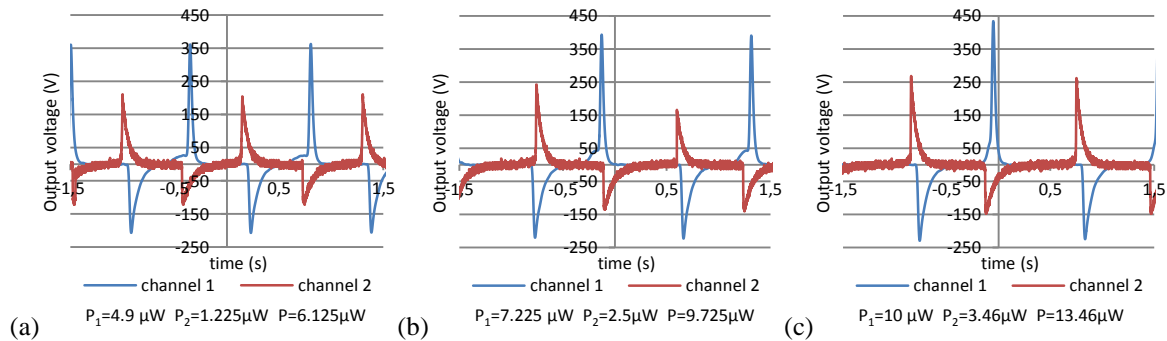

Figure 7. Output voltages and mean output powers for various electret's surface voltages (a) $V_s$=400V (b) $V_s$=450V (c) $V_s$=500V

The output voltages in open-circuit (1GΩ) are large (>400V), which is quite common in electret-based converters. As a consequence, an instantaneous output power of 200µW can be reached thanks to these devices. However, due to the low bimetal's snapping frequency (1-3Hz), we only managed to get 13.46µW of mean output power per device (for $V_s$=500V) on a 1GΩ load (channel 1 + channel 2). This corresponds to a power density of 22µW/cm³. Moreover, the snapping frequency is directly linked to the thermal gradient: the colder the upper plate, the higher the snapping frequency; and removing forced convection makes the bimetal stop. As a consequence, increasing the thermal gradient (hotter hot plate, heat sink) should increase the output power.

It is also noteworthy that (i) output voltages of channel 1 are higher than output voltages of channel 2 and (ii) the positive voltages are in both cases larger than the negative ones. The first point is due to the fact that the upper plate deforms more easily than the lower one as it is totally free to move (while the lower plate is in contact with the hotplate). Therefore, capacitances and their variations are higher in the first converter (bimetal-electret-upper plate); output voltages that are strongly linked to these two parameters are thus higher. The second point is explained by parasitic capacitances. We had already proven [19] that negative voltages are more impacted by parasitic capacitances than positive voltages; this is here again confirmed.

Finally, currently, the most important limitation of these non-optimized prototypes concerns the thermal-to-mechanical conversion, which is comprised today, for this experiment, between 0.1% and 0.5% of Carnot efficiency. Yet, we expect to increase this conversion efficiency to several per cents by optimizing the prototypes, for example by replacing air by vacuum in the cavity or by improving the bondings.

### 3.3 Lifetime

Lifetime is a critical point for energy harvesters as they are made to replace batteries and, as a consequence, they should work for at least 10 years. These devices present two elements that could greatly limit the lifetime: the bimetal and the electrets. Actually, bimetals are generally sold for 10'000 to 100'000 cycles, which corresponds to about 14 hours (in the best case) of functioning at 2Hz; ours are qualified for more than 3 million cycles. Moreover, it is well known that contacts and elevated temperatures strongly impact electrets' stability.

In order to validate the viability of this concept, two devices have been tested during 120 hours (equivalent to 5 days); this corresponds to about 850'000 cycles (at 2Hz). The electret was charged at -505V with the process mentioned above.

At the end of these 120 hours, the bimetal was still oscillating with no shift of the hysteresis cycle and the surface voltage of the electret did not drop (actually, it slightly increased to -510V, probably due to some triboelectric effects between parylene-C and Teflon). Complementary experiments are in progress to test the devices on a longer duration, but these preliminary results are extremely encouraging.

### 3.4 Pros and cons vs standard thermoelectric energy harvesters

Advantages and disadvantages of bimetal-based thermal energy harvesters compared to standard thermoelectric energy harvesters are overviewed in Table 1.

Table 1. Pros and Cons vs standard thermoelectric energy harvesters

| Advantages | Disadvantages |
|---|---|
| Low cost | AC high voltages |
| Simple and fully available materials | Low capacitances |
| High voltages | |
| Easy to manufacture and compatible with mass production | |
| Easier to keep the thermal gradient in the structure due to the cavity (limits the thermal bridge) | |
| Flexibility/Semi-flexibility | |

Due to their ease of manufacture, these devices are compatible with mass production over large areas. This can be of great interest to harvest energy on large surface environments such as walls, pipes, etc. However, this implies a power management circuit that supports non-synchronous devices in parallel.

### 4. Energy harvesting from devices in parallel and power management circuit

As presented previously, bimetal-and-electret-based thermal energy harvesters are characterized by a high output voltage that reaches some hundreds of volts and a low output current (some 100nA). Obviously, it is impossible to power any application, any electronic device with such a supply source, as a 3V DC supply source is generally required. This is the reason why a power converter and an energetic buffer are needed to develop autonomous sensors; the conversion chain that turns an energy harvester into a viable supply source for an electronic circuit is presented in figure 8. The Power Management Circuit plays an essential role in this chain and the way the power converter is controlled has a deep impact on the power extracted from the energy harvester (e.g. SSHI [21]).

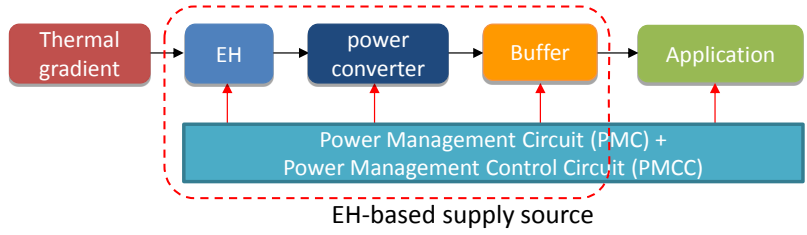

Figure 8. Conversion chain to develop viable energy harvesters

In fact, many Power Management Circuits (PMC) have been proposed in the state of the art, but, they are generally harvesting power from only one device. Yet, power transfer on energy harvesters' maximum output voltage has proven to be compatible with energy harvesting from multiple electret-based devices, as presented hereafter.

### 4.1 Power transfer on maximum voltage detection – Energy harvesting from multiple devices in parallel

Electret-based output voltages are 10 to 100 times higher than 3V: a step-down converter is therefore needed to fill the buffer. The most common step-down converters are the buck, the buck-boost and the flyback converters. Here, we have chosen to focus on the flyback converter which is simple, and therefore low consumptive to control, as only 2 controlled transistors are required.

The concept of "power transfer on maximum voltage detection" is to send the energy from the energy harvesters to the buffer ($C_b$) when EHs' output voltage reaches its maximum (figure 9(a)). The Power Management Control Circuit (PMCC), which controls the transistors, is aimed at finding the maximum voltage across the energy harvester. Then, the PMCC closes $K_p$ to transfer the energy from the energy harvesters to the magnetic circuit and closes $K_s$ to send the energy from the magnetic circuit to the buffer $C_b$.

This power management circuit is particularly suitable for one energy harvester (figure 9(a)), but can be adapted to multiple devices in parallel as presented in figure 9(b) for 2 devices. In this configuration, power is transferred from the energy harvesters in parallel to the storage element as soon as one of the energy harvesters' output voltage reaches its maximum.

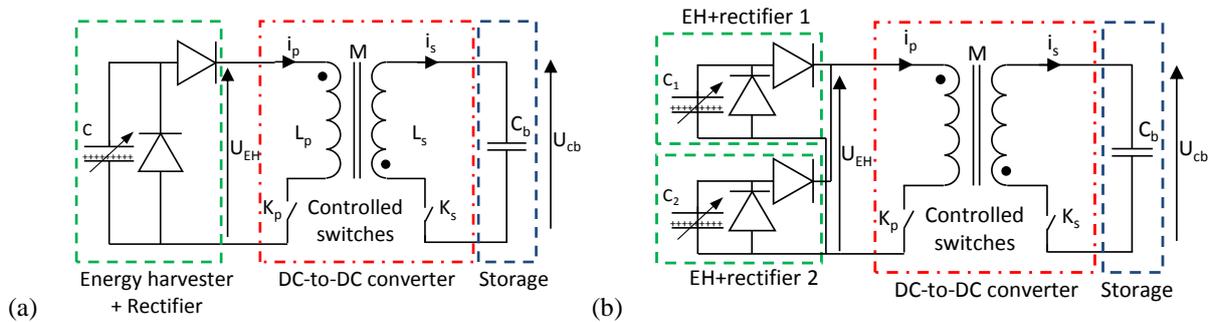

Figure 9. Power Management Circuit – power transfer on energy harvesters' maximum voltage (a) for one device and (b) for 2 devices in parallel

Theoretical voltages ($U_{EH}$, $U_{cb}$) and currents on the primary ($i_p$) and on the secondary ($i_s$) windings during the power transfer in the flyback converter are presented in figure 10. As soon as $U_{EH}$ reaches its maximum, $K_p$ is closed (t=$t_0$). The primary (C, $L_p$) behaves like a LC circuit. Then, the energy stored into the capacitance C of the energy harvester is totally transferred to the primary inductance $L_p$ and stored into the magnetic circuit in one quarter of a period of the $L_p$-C circuit, which corresponds to $T_1$. As a consequence, $U_{EH}$ drops to 0 and the current $i_p$ increases up to $I_{pmax}$ at $t_0+T_1$. $U_{cb}$ stays constant and equal to $U_{cb}^-$ ($U_{cb}$ before energy transfer). $T_2$ is a guard time, common in flyback converters. Then, $K_s$ is closed. The secondary ($L_s$-$C_b$) behaves like a LC circuit. The energy stored into the magnetic circuit is transferred to $L_s$ and finally to $C_b$. Then, $U_{cb}$ increases from $U_{cb}^-$ to $U_{cb}^+$. Thanks to this circuit, the most part of the energy stored into the capacitance of the energy harvester is transferred to the storage capacitor $C_b$.

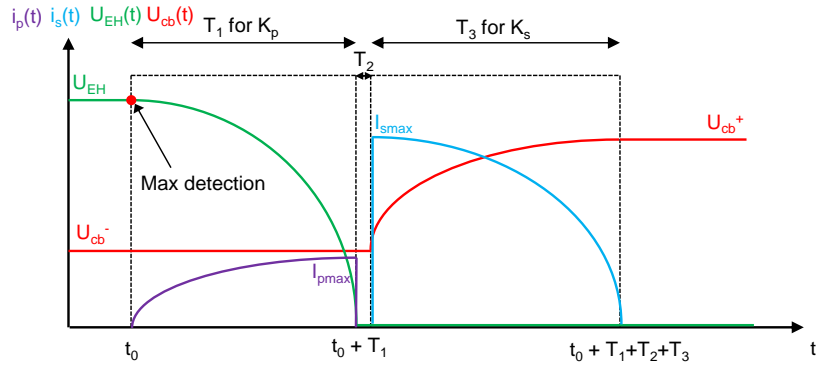

Figure 10. Flyback DC-to-DC converter – theoretical voltages and currents during power conversion

The power conversion using a flyback converter can reach 70-80% of efficiency, assuming that it is well-designed; and actually, the operating frequency of the flyback, the material of the magnetic core, the number of windings and the transistors, which have a major impact on the converter's efficiency, have been specifically and carefully chosen for this application. Table 2 overviews the flyback's parameters that have been taken to maximize the conversion efficiency.

Table 2. Flyback's parameters

| Parameter | Value |
| --- | --- |
| Operating frequency | 100kHz |
| Core Material | 3F3 |
| Air gap | 0μm |
| Windings on the primary | 320 |
| Windings on the secondary | 10 |
| $L_p$ | 98.2mH (experimental data) |
| $L_s$ | 85μH (experimental data) |

As explained previously, to control the PMC (and especially $K_p$ and $K_s$), a power management control circuit (PMCC) is required. Its principle and its constitutive functions are introduced in figure 11(a): energy harvesters' output voltage is derived by a RC derivator ($C_D$-$R_D$), which is then compared to 0 with a MAX919 comparator, known as a low consumption IC. These two first steps enable to detect when the energy harvesters' output voltage reaches its maximum. Then, three delay cells, made from simple logic components (buffer, inverter, AND gate) generate the 2 control times ($T_1$ for $K_p$ and $T_3$ for $K_s$) when the comparator switches to its 'high' state. Control times for $K_p$ and $K_s$ are recapitulated in figure 11(b), and here, $T_1$ is equal to ~5μs, $T_2$ to ~1μs and $T_3$ to ~10μs.

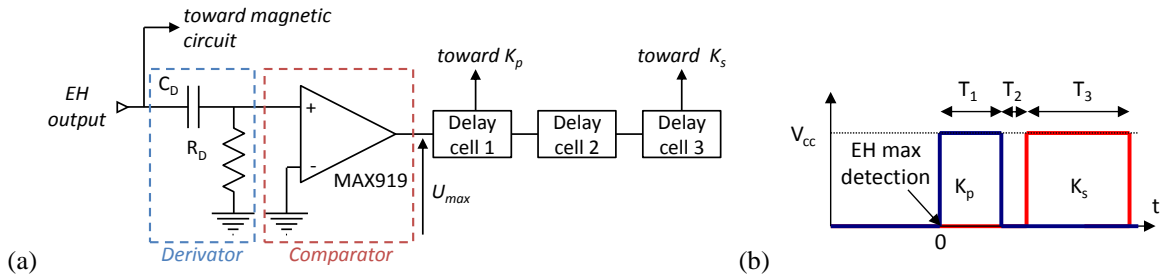

Figure 11. Power Management Control Circuit (PMCC) (a) functions and (b) control times for $K_p$ and $K_s$

The drawback of using a PMCC is the inherent power consumption of electronic components that detect the maximum voltage and generate the control times. Yet, as based on low-consumption integrated circuits (simple logic components, MAX919), our PMCC consumes only 500nA@3V.

### 4.2 Experimental results on the Power Management Circuit

The maximum voltage detection using the PMC has been validated on 3 devices. Figure 12(a) shows the output voltage $U_{EH}$ with three devices in parallel on a 1GΩ load and without the power conversion step, and the maximum voltage detection ($U_{max}$) performed by the PMCC, showing the proper functioning of the circuit. This is confirmed when the power conversion step is added (figure 12(b)). But obviously, maximum voltage detection now leads to a power transfer from the energy harvesters' capacitors to the buffer capacitor $C_b$: $U_{EH}$ drops to 0 after the maximum voltage is detected. It is also noteworthy that voltage peaks of $U_{EH}$ are reduced when the power converter is added (figure 12b vs figure 12a). Actually, this is due to parasitic capacitances induced by the windings and the transistors. They strongly reduce the voltage peaks and as a consequence the output power.

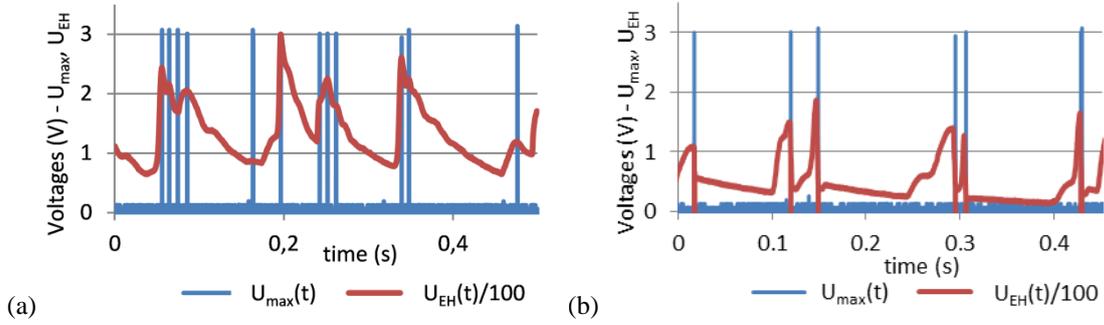

Figure 12. Maximum voltage detection for 3 devices in parallel (a) with no power conversion step and (b) with the power conversion step

The power conversion with the flyback has also been validated (figure 13). The maximum voltage detection occurs at t=0: $K_p$ is closed, the energy stored into the capacitors of the energy harvesters is sent to the magnetic circuit, leading to a drop of energy harvesters' output voltage to 0. Then, $K_s$ is closed, and the energy stored into the magnetic circuit is transferred to $C_b$; $i_s$ circulates on the secondary winding, reaching about 400mA, and $U_{cb}$, the voltage across $C_b$, increases. The DC-to-DC power conversion reaches about 70% of efficiency.

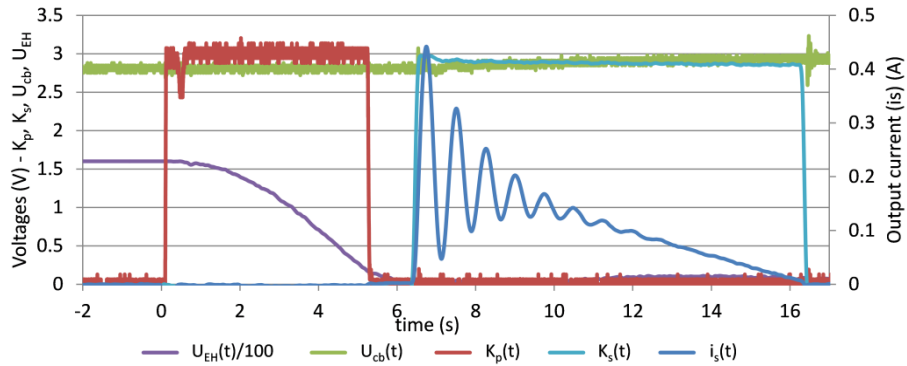

Figure 13. Current $i_s$ and voltages $U_{EH}$, $U_{cb}$, $K_p$, $K_s$ during power conversion with the flyback DC-to-DC converter

Finally, thanks to 3 devices, placed on a hot source at 60°C, cold by forced convection and with $V_s$=500V (like in section 3), we managed to get about 10µW of directly usable output power (5.45µW/cm³): 5.6V have been stored in a 230µF capacitor in 350s (figure 14(a)); this has been confirmed by a voltage measurement in the steady state on a load of 1MΩ in parallel with a 10µF capacitor placed at the output of the magnetic circuit (figure 14(b)).

Yet, 10µW represent only 25% of the sum of the 3 energy harvesters' output powers (3×13.46µW=40.38µW). In fact, this is due to the introduction of parasitic capacitances, which strongly affect electret-based energy harvesters' output powers, by:

(i) the parallelization of the energy harvesters. Actually, the capacitance of an energy harvester is perceived as a parasitic capacitance by the other energy harvesters.

(ii) the flyback converter, which introduces high parasitic capacitances in parallel with the energy harvesters.

Anyway, 10μW are compatible with a WSN powering application, providing that an intermittent running mode is adopted [18].

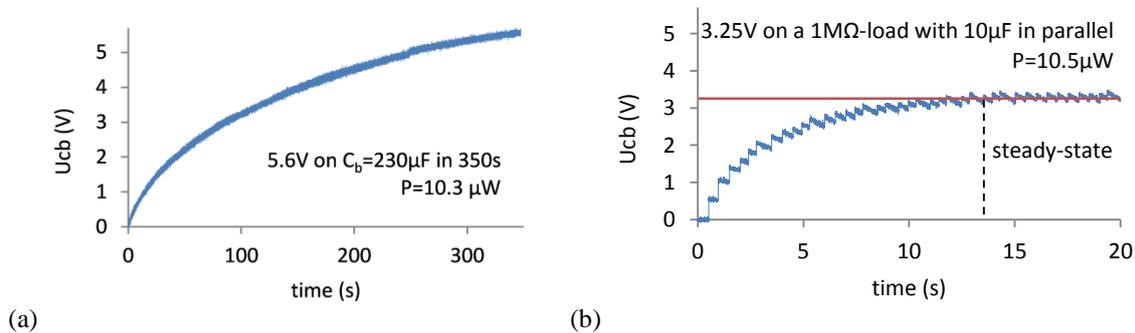

Figure 14. (a) Output voltage on a 230μF buffer capacitor as a function of the time and (b) output voltage on a 1 MΩ load in parallel with a 10μF buffer capacitor.

## 5. Conclusions and Perspectives

We have presented semi-flexible devices able to harvest power from thermal gradients by using a two-steps conversion and based on a bimetal and an electret-based converter. We managed to get a mean output power of 13.46μW per device on a hot source at 60°C and forced convection. A power management circuit that implements a "power transfer on maximum voltage detection" using a flyback DC-to-DC power converter has been developed and tested on 3 devices in parallel. 10μW of usable output power were obtained with these 3 devices, which is enough to power a Wireless Sensor Node with an intermittent running mode.

Yet, the strong impact of parasitic capacitances induced by the flyback converter has been shown: research is now focused on a way to reduce them by improving windings and electronic components. Similarly, investigations are carried out to remove the forced convection: working at higher temperatures or increasing the cavity size are two means to perform this. Moreover, limiting the conduction from the lower plate to the bimetal and to the upper plate through the bondings is also under investigation.

These new devices pave the way to low-cost thermal energy harvesters manufacturable over large areas, and especially suitable for energy harvesting on large surface environments such as walls, pipes, roofs of cars, etc. The semi-flexible property is a plus that allows to adapt bimetal-based thermal energy harvesters to any kind of surface.

**Acknowledgments**

The authors would like to thank G. Pitone and G. Delepierre, DeltaConcept, for the design of the bimetals.

This work was partially funded by the European Regional Development Fund (FEDER/ERDF) and by the French Inter-ministerial Fund (FUI), through HEATec project.